\documentclass{JHEP3}

  \usepackage{latexsym,bm,amsmath,amssymb,amsfonts}
  \usepackage{epsfig,graphics,graphicx}
  \usepackage{slashed}
  \usepackage[latin1]{inputenc}
  \usepackage{mathrsfs}
\usepackage{mathcomp}

  \long\def\comment#1{ }

\def\be{\begin{equation}}
\def\ee{\end{equation}}
\def\bea{\begin{eqnarray}}
\def\eea{\end{eqnarray}}

\title{\rm \LARGE \bf Back reaction, emission spectrum and entropy spectroscopy}

\author{Qing-Quan Jiang$^{a,b}$ and Xu Cai$^{b}$\\

$^{a}$ Institute of Theoretical Physics, China
 West Normal University, Nanchong, Sichuan \\637002, People's Republic of
China\\
$^{b}$ Institute of Particle Physics, Central China Normal
University, Wuhan, Hubei 430079, \\People's Republic of China\\
{\tt E-mail address: jiangqq@iopp.ccnu.edu.cn, xcai@mail.ccnu.edu.cn}}

\abstract{Recently, an interesting work, which reformulates the tunneling framework to directly produce the Hawking emission spectrum and entropy spectroscopy in the tunneling picture, has been received a broad attention. However, during the emission process, most related observations have not incorporated the effects of back reaction on the background spacetime, whose derivations are therefore not the desiring results for the real physical process. With this point as a central motivation, in this paper we suitably adapt the \emph{reformulated} tunneling framework so that it can well accommodate the effects of back reaction to produce the Hawking emission spectrum and entropy spectroscopy. Consequently, we interestingly find that, when back reaction is considered, the Parikh-Wilczek's outstanding observations that, an isolated radiating black hole has an unitary-evolving emission spectrum that is \emph{not} precisely thermal, but is related to the change of the Bekenstein-Hawking entropy, can also be reproduced in the reformulated tunneling framework, meanwhile the entropy spectrum has the same form as that without inclusion of back reaction, which demonstrates the entropy quantum is \emph{independent} of the effects of back reaction. As our final analysis, we concentrate on the issues of the black hole information, but \emph{unfortunately} find that, even including the effects of back reaction and higher-order quantum corrections, such tunneling formalism can still not provide a mechanism for preserving the black hole information.}

\keywords{Black holes, Modes of Quantum Gravity}

\begin{document}

\section{Introduction}

In 1974, Hawking proved that black hole can radiate particles from
its event horizon with a temperature proportional to its surface
gravity, and the emission spectrum is a purely thermal one
\cite{r1}. Since then, a lot of attempts have been appeared to reproduce the Hawking radiation from black hole. Among them, an intuitively simple but physically rich framework, which treats the Hawking radiation as a semiclassical tunneling process at the horizon of black hole, has been received a popular attention  \cite{PW,PW1,PW2,NPW,PW3,PW4,PW5,PW6,PW7,PW8}. The semiclassical Hawking temperature is very simply and quickly obtained in this scheme by exploiting the form of the semiclassical tunneling rate. Unfortunately, this tunneling framework fails to yield directly the Hawking emission spectrum. With
this shortcoming as a central motivation, Banerjee and Majhi, with the aid of density matrix techniques, have successfully reformulated the tunneling framework so that it can well describe the emission spectrum from black hole \cite{c1,c2}. Later on, the entropy spectrum was well described in the \emph{reformulated} tunneling analysis \cite{BRM,RBE,JHC}. In view of these developments, the tunneling picture appears more robust and convincing to describe the emission behavior of black hole.

There has been a great deal of success at using this \emph{reformulated} tunneling framework as a means of reproducing the emission spectrum and entropy spectroscopy of black hole \cite{c1,c2,BRM,RBE,JHC}. However, most relevant works neglected the effects of back reaction on the background spacetime; opting to work, as a matter of choice, with the classical geometry throughout. In fact, when a particle is emitted from the horizon, the energy conservation demands that the background geometry may transit from one state to another with a lower black hole mass (This supports the idea
that, in quantum gravity, black holes are properly regarded as highly excited states). Which is to say, the background geometry should be considered in dynamical transition during the emission process \cite{PW,PW1,PW2,NPW,PW3,PW4}.
Hence, the previous treatments on this topic \cite{c1,c2,BRM,RBE,JHC}, with a fixed background spacetime, were a first try to face the
problem of black hole emission spectrum and now we have decided to
add one more element, i.e. the back reaction effect, that makes the problem
more physical but also more difficult.
Meanwhile, many earlier works have stressed the important significance of the tunneling paradigm for incorporating the effects of back-reaction on the background
spacetime; that is, the blackbody spectrum and Bekenstein-Hawking entropy with inclusion of back reaction would be so modified that an isolated radiating black hole has an unitary-evolving emission spectrum that is not precisely thermal, but is related to the change of the Bekenstein-Hawking entropy \cite{PW,PW1,PW2,NPW,PW3,PW4}. Armed with these insights, a quite natural generalization of the \emph{reformulated} tunneling analysis would be to see if its formalism can be viably extended into this quantum-gravitational regime, and in what follows, to observe what important significance happens to the emission spectrum and entropy spectroscopy due to the effects of back reaction. Such an extension is, in fact, the objective of the current paper.

In this paper, we attempt to develop the \emph{reformulated} tunneling analysis for accommodating the effects of back reaction $-$ namely, in the reformulated tunneling framework producing the Hawking emission spectrum and entropy spectroscopy with back reaction. Our result shows that, in the \emph{reformulated} tunneling framework, the Hawking emission spectrum with back reaction is indeed not precisely thermal, but is related to the change of the Bekenstein-Hawking entropy, as described in the Parikh-Wilczek's tunneling framework; Meanwhile the entropy spectrum has the form $S_n=n$ as that without inclusion of back reaction. This is an interesting result since the non-thermal Hawking spectrum with back reaction implies an underlying unitary theory, which may provide a possible explanation to the long-standing ``loss of information" (In fact, our final analysis shows, even including the effect of back reaction and higher-order quantum corrections, such tunneling formalism can still not provide a mechanism for preserving the black hole information.);  In addition, the entropy quantum is independent of the effects of back reaction, which is an intriguing result for quantum-gravitational principles.

The remainders of this paper are outlined as follows. In
Sec.\ref{sec1}, we briefly recall the main results of the reformulated tunneling framework developed by Banerjee and Majhi. Then, developing the formulated tunneling analysis for incorporating the effects of back reaction, Sec.\ref{sec2} and \ref{sec3} are, respectively, devoted to producing the Hawking emission spectrum and entropy spectroscopy with back reaction. Sec.\ref{sec4} ends up with
some discussions and conclusions on our results.

\section{Emission spectrum and entropy spectroscopy via tunneling} \label{sec1}

In this section, we simply review the reformulated tunneling method to obtain the Hawking emission spectrum and entropy spectroscopy from the event horizon of black hole, whose motivation is to compare their results with those with inclusion of back reaction. At the black hole horizon, when particles are created just inside the horizon, the right moving mode is tunneling across it to be observed by the observer living outside, while the left moving mode is trapped inside the horizon. Then, a question arises how to connect the right and left moving modes inside and outside the horizon. In the Boyer-Lindquist coordinate system, the fact that the coordinate singularity appears at the horizon means the metric can only be defined either inside or outside the horizon, in this case, such connection appears difficult. To connect the modes outside and inside the horizon, we should first find a coordinate system in which the metric can be effectively defined both inside and outside the horizom. In fact, the Kruskal extension is done by concentrating on the behavior at the horizon only. In this case, proceeding in a similar way in Refs. \cite{c1,c2,BRM,RBE,JHC}, one can find the right and left modes inside and outside the horizon are related by
\begin{eqnarray}
\Phi_{\textrm{in}}^L&=&\Phi_{\textrm{out}}^L \nonumber\\
\Phi_{\textrm{in}}^R&=&e^{-\frac{\pi E}{\hbar \kappa_h}}\Phi_{\textrm{out}}^R.\label{eq1}
\end{eqnarray}
Here, $E$ is the effective energy of the emitted particle detected by the observer living outside the horizon, and $\kappa_h$ is the surface gravity at the horizon of the black hole. In the following subsections, with the aid of the connection between the right and left moving modes inside and outside the horizon, we will produce the Hawking emission spectrum and entropy spectroscopy from the horizon of the black hole.

\subsection{Quantum tunneling and emission spectrum} \label{sec2.1}

In this subsection, with the aid of density matrix technique, we will present the Hawking emission spectrum in the reformulated tunneling framework developed in \cite{c1,c2}. When $n$ number of non-interacting virtual pairs are produced just inside the horizon, each pair defined by the left side of (\ref{eq1}), any physical state of the system, detected by an observer outside the horizon, is given by \cite{c1,c2}
\begin{equation}
|\Psi\rangle=N\sum_n|n_{\textrm{in}}^L\rangle\otimes |n_{\textrm{in}}^R\rangle=N\sum_n e^{-\frac{\pi n E}{\hbar \kappa_h}}|n_{\textrm{out}}^L\rangle\otimes |n_{\textrm{out}}^R\rangle,\label{eq2}
\end{equation}
where we have used the relation (\ref{eq1}) between the left and right modes inside and outside the horizon, and $N$ is a normalization constant, which can be determined by the normalization condition $\langle\Psi|\Psi\rangle=1$. This immediately yields $N=(\sum_n e^{-\frac{2\pi n E}{\hbar \kappa_h}})^{-\frac{1}{2}}$. For bosons, where $n=0,1,2,3,\cdot\cdot\cdot\cdot\cdot\cdot$, the sum will be calculated as $N_{\textrm{boson}}=(1-e^{-\frac{2\pi E}{\hbar \kappa_h}})^{\frac{1}{2}}$. For fermions, where $n=0,1$, the sum becomes $N_{\textrm{fermion}}=(1+e^{-\frac{2\pi E}{\hbar \kappa_h}})^{-\frac{1}{2}}$. In our following analysis, we only consider the case of bosons without loss of generality, since for fermions the analysis is identical. For bosons, the density matrix can be constructed as
\begin{eqnarray}
\widehat{\rho}_{\textrm{boson}}&=&|\Psi\rangle_{\textrm{boson}}\langle\Psi|_{\textrm{boson}}\nonumber\\
&=& (1-e^{-\frac{2\pi E}{\hbar \kappa_h}})\sum_{n,m} e^{-\frac{\pi n E}{\hbar \kappa_h}}e^{-\frac{\pi m E}{\hbar \kappa_h}}|n_{\textrm{out}}^L\rangle\otimes |n_{\textrm{out}}^R\rangle \langle m_{\textrm{out}}^R|\otimes \langle m_{\textrm{out}}^L|. \label{eq3}
\end{eqnarray}
Now, tracing out the left modes, which the observer living outside the horizon can not detect, the reduced density operator for the right modes is written as
\begin{equation}
\widehat{\rho}_{\textrm{boson}}^R=(1-e^{-\frac{2\pi E}{\hbar \kappa_h}})\sum_n e^{-\frac{2 \pi n E}{\hbar \kappa_h}}|n_{\textrm{out}}^R\rangle \langle n_{\textrm{out}}^R|.\label{eq4}
\end{equation}
Then, the average number of particles detected at asymptotic infinity is given by
\begin{equation}
\langle n\rangle_{\textrm{boson}}=\textrm{trace}(\widehat{n}\widehat{\rho}_{\textrm{boson}}^R)=({e^{\frac{2\pi E}{\hbar \kappa_h}}-1})^{-1}.\label{eq5}
\end{equation}
This is nothing but the Bose-Einstein distribution of particles, which corresponds to the blackbody spectrum with the temperature $T_h=\frac{\hbar\kappa_h}{2\pi}$. In a similar way, the Hawking emission spectrum for fermions is
\begin{equation}
\langle n\rangle_{\textrm{fermion}}=\textrm{trace}(\widehat{n}\widehat{\rho}_{\textrm{fermion}}^R)=({e^{\frac{2\pi E}{\hbar \kappa_h}}+1})^{-1}.\label{eq6}
\end{equation}
In short, with the help of density operator, we can produce the Hawking emission spectrum for bosons and fermions in the tunneling framework. Note also that the Hawking emission spectrum (\ref{eq5}) and (\ref{eq6}) for bosons and fermions are currently presented as a precisely thermal spectrum. In such case, the information carried
by a physical system falling toward black hole singularity
has no way to be recovered after a black hole has disappeared
completely, since the radiation with a precisely thermal
spectrum carries no information. This is the so-called ``information loss
paradox", which means that pure quantum states (the
original matter that forms the black hole) can evolve into
mixed states (the thermal spectrum at infinity). Such an
evolution violates the fundamental principles of quantum
theory, as these prescribe a unitary time evolution of basis
states. With inclusion of back reaction, Sec. \ref{sec2} will present an unitary-evolving emission
spectrum that is not precisely thermal, but is related to the change of the Bekenstein-Hawking entropy.
In the next subsection, we will see how to describe the entropy spectrum in the reformulated tunneling method.

\subsection{Quantum tunneling and entropy spectroscopy}

In this subsection, we focus on studying the entropy spectrum in the reformulated tunneling framework developed in \cite{BRM,RBE,JHC}. As mentioned above, the left and right moving modes inside and outside the horizon are connected by (\ref{eq1}). Also, the left moving modes are trapped inside the horizon, while the right moving modes can tunnel across the horizon to be observed at asymptotic infinity. In this case, the average value of the energy $E$ can be written as
\begin{eqnarray}
\langle E\rangle=\frac{\int_0^\infty (\Phi_{\textrm{in}}^R)^\ast E \Phi_{\textrm{in}}^R dE }{\int_0^\infty (\Phi_{\textrm{in}}^R)^\ast  \Phi_{\textrm{in}}^R dE}&=&\frac{\int_0^\infty e^{-\frac{\pi E}{\hbar \kappa_h}}(\Phi_{\textrm{out}}^R)^\ast E e^{-\frac{\pi E}{\hbar \kappa_h}}\Phi_{\textrm{out}}^R dE }{\int_0^\infty e^{-\frac{\pi E}{\hbar \kappa_h}}(\Phi_{\textrm{out}}^R)^\ast  e^{-\frac{\pi E}{\hbar \kappa_h}}\Phi_{\textrm{out}}^R dE}\nonumber\\
&=& \frac{\int_0^\infty E e^{-\frac{2\pi E}{\hbar \kappa_h}}dE}{\int_0^\infty  e^{-\frac{2\pi E}{\hbar \kappa_h}}dE}=\frac{\hbar\kappa_h}{2\pi}=T_h, \label{eq7}
\end{eqnarray}
where $T_h=\frac{\hbar\kappa_h}{2\pi}$ is identified as the Hawking temperature at the horizon of the black hole. In a similar way, we can compute the average squared energy of particles detected by the asymptotic observer as
\begin{eqnarray}
\langle E^2\rangle=\frac{\int_0^\infty (\Phi_{\textrm{in}}^R)^\ast E^2 \Phi_{\textrm{in}}^R dE }{\int_0^\infty (\Phi_{\textrm{in}}^R)^\ast  \Phi_{\textrm{in}}^R dE}&=&\frac{\int_0^\infty e^{-\frac{\pi E}{\hbar \kappa_h}}(\Phi_{\textrm{out}}^R)^\ast E^2 e^{-\frac{\pi E}{\hbar \kappa_h}}\Phi_{\textrm{out}}^R dE }{\int_0^\infty e^{-\frac{\pi E}{\hbar \kappa_h}}(\Phi_{\textrm{out}}^R)^\ast  e^{-\frac{\pi E}{\hbar \kappa_h}}\Phi_{\textrm{out}}^R dE}\nonumber\\
&=& \frac{\int_0^\infty E^2 e^{-\frac{2\pi E}{\hbar \kappa_h}}dE}{\int_0^\infty  e^{-\frac{2\pi E}{\hbar \kappa_h}}dE}=\frac{\hbar^2\kappa_h^2}{2\pi^2}=2T_h^2. \label{eq8}
\end{eqnarray}
Now, combining the equations (\ref{eq7}) and (\ref{eq8}), one can easily find the uncertainty in the detected energy $E$ is given by
\begin{equation}
\Delta E=\sqrt{\langle E^2\rangle-\langle E\rangle^2}=T_h.\label{eq9}
\end{equation}
This uncertainty can be treated as the lack of information in energy of the black hole during the particle emission. Also, in the information theory,
the entropy acts as the lack of information. To connect these two quantities, we can use the first law of thermodynamic as $\Delta E=T_h\Delta S_{\textrm{bh}}$. In view of (\ref{eq9}), the uncertainty in the characteristic frequency of the tunneling mode is $\Delta f=\frac{\Delta E}{\hbar}=\frac{T_h}{\hbar}$. According to the Bohr-Sommerfeld quantization rule, substituting this uncertainty into the first law of thermodynamic yields the entropy spectrum as
\begin{equation}
 S_{\textrm{bh}}=n,\label{eq10}
\end{equation}
where $n$ is an integer. From Eq.(\ref{eq10}), $\Delta S_{\textrm{bh}}=(n+1)-n=1$, this result shows that the entropy of the black hole is quantized in units of the identity. Also, the similar nature of the entropy spectrum was reproduced in Refs. \cite{NEW1,NEW2} by using other different methods. Obviously, the entropy spectrum of the black hole can also be described in the tunneling framework. Till now, we have successfully described the Hawking emission spectrum and entropy spectroscopy in the reformulated tunneling framework, however, during the emission process, the effects of back reaction are not included in this analysis. Therefore, the results (\ref{eq5}), (\ref{eq6}) and (\ref{eq10}) are not the desiring results for the real physical process. In what follows, by developing the reformulated tunneling analysis for incorporating the effects of back reaction on the background spacetime, we concentrate on obtaining the Hawking emission spectrum and entropy spectroscopy with inclusion of back reaction.

\section{Back reaction and emission spectrum} \label{sec2}

As mentioned above, the reformulated tunneling framework developed in \cite{c1,c2,BRM,RBE,JHC} purposefully neglects the back reaction of
the particle on the background spacetime. So one might well ask as to how such an effect
can then be incorporated. To address this query, the sensible presumption of energy conservation has a very important implication \cite{PW,PW1,PW2,NPW,PW3,PW4}. In such case,
when a particle with the energy $E$ tunnels out across the horizon, it seems that the equation (\ref{eq1}) would be rewritten with a simple replacement $\kappa_h (M)\rightarrow \kappa_h (M-E)$ due to energy conservation. However, this treatment appears inappropriate, since the quantum
uncertainty principle demands it is unnatural to expect that the black hole mass can jump, from
$M$ to $M - E$, in such a discontinuous manner. Rather, quantum blurring will require
a gradual transition\footnote{This is relative to whatever time scale is characteristic of the radiation
process.}, so that it is much more accurate to replace $\kappa_h (M)$ with $\kappa_h (M-\omega)$ and then
suitably integrate over $\omega$. So, in the presence of back reaction, the equation (\ref{eq1}) can then be rewritten as
\begin{eqnarray}
\Phi_{\textrm{in}}^L&=&\Phi_{\textrm{out}}^L, \nonumber\\
\Phi_{\textrm{in}}^R&=&e^{-\frac{1}{2}\int_0^E \beta(M-\omega)d\omega}\Phi_{\textrm{out}}^R.\label{eq11}
\end{eqnarray}
Here, $\beta(M-\omega)=\frac{2\pi }{\hbar \kappa_h(M-\omega)}$ is the inverse of the Hawking temperature after emitted a particle with the energy $\omega$.
Now, our focus is on calculating the integration $\int_0^E \beta(M-\omega)d\omega$. Since the black hole has much more energy than the emitted particle ($M\gg\omega$), Taylor expanding in $\omega$ then yields
\begin{equation}
\int_0^E \beta(M-\omega)d\omega=\int_0^E\Big[\beta(M)-\omega\partial_M\beta(M)+\omega^2\frac{\partial_M^2\beta(M)}{2!}+\mathcal{O}(\omega^3)\Big]d\omega. \label{eq12}
\end{equation}
Now, integrating equation (\ref{eq12}), we have
\begin{equation}
\int_0^E \beta(M-\omega)d\omega=\beta(M)E-E^2\frac{\partial_M\beta(M)}{2!}+E^3\frac{\partial_M^2\beta(M)}{3!}+\mathcal{O}(E^4). \label{eq13}
\end{equation}
Recalling the first law of thermodynamic $\beta(M)=\partial_M S(M)$, where $S(M)$ is the Bekenstein-Hawking entropy of the black hole, the equation (\ref{eq13}) can then be rewritten as
\begin{eqnarray}
\int_0^E \beta(M-\omega)d\omega&=&E\partial_M S(M)-E^2\frac{\partial_M^2S(M)}{2!}+E^3\frac{\partial_M^3S(M)}{3!}+\mathcal{O}(E^4)\nonumber\\
&=& -[S(M-E)-S(M)]=-\Delta S_{\textrm{BH}},\label{eq14}
\end{eqnarray}
with $\Delta S_{\textrm{BH}}=S(M-E)-S(M)$ representing the change of the Bekenstein-Hawking entropy before and after the particle emission. Thus, in the presence of back reaction, the left and right moving modes defined inside and outside the horizon are connected by
\begin{eqnarray}
\Phi_{\textrm{in}}^L&=&\Phi_{\textrm{out}}^L, \nonumber\\
\Phi_{\textrm{in}}^R&=&e^{\frac{1}{2}\Delta S_{\textrm{BH}}}\Phi_{\textrm{out}}^R.\label{eq15}
\end{eqnarray}
In view of this connection (\ref{eq15}), one can proceed in a similar way as Sec. \ref{sec2.1} to produce the Hawking emission spectrum with back reaction. For the moment when the particles are created just inside the horizon, each defined by the modes in the left side of (\ref{eq15}), the physical state (\ref{eq2}) for the $n$ number of non-interacting virtual pairs, in the presence of back reaction, can then be rewritten as
\begin{equation}
|\Psi\rangle=N\sum_n|n_{\textrm{in}}^L\rangle\otimes |n_{\textrm{in}}^R\rangle=N\sum_n e^{\frac{1}{2}n\Delta S_{\textrm{BH}}}|n_{\textrm{out}}^L\rangle\otimes |n_{\textrm{out}}^R\rangle,\label{eq16}
\end{equation}
where $|n_{\textrm{out}}^L\rangle$ and $|n_{\textrm{out}}^R\rangle$ respectively correspond to the $n$ number of left and right moving modes, and $N$ is a normalization constant, whose value can be determined by the normalization condition $\langle\Psi|\Psi\rangle=1$, which immediately yields $N=(\sum_ne^{n\Delta S_{\textrm{BH}}})^{-\frac{1}{2}}$. For bosons, $n=0,1,2,3\cdot\cdot\cdot\cdot\cdot\cdot$, the sum is calculated as
\begin{equation}
N_{\textrm{boson}}=(1-e^{\Delta S_{\textrm{BH}}})^{\frac{1}{2}}.
\end{equation}
While, for fermions, where $n=0,1$, the sum becomes
\begin{equation}
N_{\textrm{fermion}}=(1+e^{\Delta S_{\textrm{BH}}})^{-\frac{1}{2}}.
\end{equation}
In our case, we only consider the case of bosons without loss of generality since the analysis is identical for fermions. In view of the physical state (\ref{eq16}), the density matrix operator for bosons in the presence of back reaction is given by
\begin{eqnarray}
\widehat{\rho}_{\textrm{boson}}&=&|\Psi\rangle_{\textrm{boson}}\langle\Psi|_{\textrm{boson}}\nonumber\\
&=& (1-e^{\Delta S_{\textrm{BH}}})\sum_{n,m} e^{\frac{1}{2}n\Delta S_{\textrm{BH}}}e^{\frac{1}{2}m\Delta S_{\textrm{BH}}}|n_{\textrm{out}}^L\rangle\otimes |n_{\textrm{out}}^R\rangle \langle m_{\textrm{out}}^R|\otimes \langle m_{\textrm{out}}^L|. \label{eq17}
\end{eqnarray}
Now, tracing out the left moving modes, which cannot be detected by the observer living outside, the reduced density matrix for the right moving modes is read off to be
\begin{equation}
\widehat{\rho}_{\textrm{boson}}^R=(1-e^{\Delta S_{\textrm{BH}}})\sum_n e^{n\Delta S_{\textrm{BH}}}|n_{\textrm{out}}^R\rangle \langle n_{\textrm{out}}^R|.\label{eq18}
\end{equation}
Then, based on the density matrix (\ref{eq18}), the average number of particles detected at asymptotic infinity is given by
\begin{equation}
\langle n\rangle_{\textrm{boson}}=\textrm{trace}(\widehat{n}\widehat{\rho}_{\textrm{boson}}^R)=({e^{-\Delta S_{\textrm{BH}}}-1})^{-1}.\label{eq19}
\end{equation}
This is the Hawking emission spectrum for bosons with back reaction, which is nothing but the Bose-Einstein distribution with the effective (inverse) temperature \begin{equation}
\beta_{\textrm{eff}}=\beta(M)-E\frac{\partial_M\beta(M)}{2!}+E^2\frac{\partial_M^2\beta(M)}{3!}+\mathcal{O}(E^3),\label{eq20}
\end{equation}
where we have used the relations between (\ref{eq13}) and (\ref{eq14}). As a result, it is found that, in the presence of back reaction, the Hawking emission spectrum for bosons deviates from the precisely thermal one \footnote{This can be deduced form the energy-dependent effective temperature (\ref{eq20}).}, but is related to the change of the Bekenstein-Hawking entropy before and after the particle emission. Whereas for fermions, in a similar way, the Hawking emission spectrum in the presence of back reaction is given by
\begin{equation}
\langle n\rangle_{\textrm{fermion}}=\textrm{trace}(\widehat{n}\widehat{\rho}_{\textrm{fermion}}^R)=({e^{-\Delta S_{\textrm{BH}}}+1})^{-1}.\label{eq21}
\end{equation}
This is the Fermi-Dirac distribution with the effective temperature (\ref{eq20}). Apparently, in the presence of back reaction, the Hawking emission spectrum deviates from the precisely thermal one (\ref{eq5}) and (\ref{eq6}), but is related to the difference of the Bekenstein-Hawking entropy before and after the particle emission. This result implies a unitary theory underlying the process of black hole evaporation (appears in Appendix A), which is expected from the viewpoint of quantum theory. Meanwhile, our observation is in agreement with other treatments \cite{PW,PW1,PW2,NPW,PW3,PW4}, and therefore substantiates our ansatz for incorporating the effects of back reaction in the reformulated tunneling framework developed in \cite{c1,c2}. Note that, in spite of the non-thermal spectrum and implied unitary process for black hole emission, it is not at all clear as to how the information might actually be preserved. Considering the emission of two particles $E_1$ and $E_2$, and the emission of one particle with their combined energies $E_1 +E_2$, we find from the equation (\ref{eqA3}), $\ln[P(E_1)P(E_2)]=\ln[P(E_1+E_2)]$. This means that the probability of emitting two particles of energy $E_1$ and $E_2$ is
exactly the same as the probability of emitting just a single particle of the same total
energy, so there is no correlation, at least at late-times \footnote{The tunneling methodology can not currently address the
issue of short-time correlations.} \cite{NPW}. That is to say, the form of this correction is
not sufficient by itself to recover information. Recently, it was shown that under specific conditions if one takes into account both the
back reaction effect and the quantum gravity corrections to all order of $\hbar$, information 
can be retrieved when the black hole is completely evaporated
with the information carried away by the correlations of the outgoing radiation \cite{NEW3}.

\section{Back reaction and entropy spectroscopy } \label{sec3}

In this section, with the aid of the reformulated tunneling framework, we will present the entropy spectrum with back reaction. We have shown in the previous section that pair production occurs inside the horizon, and the left and right moving modes defined inside and outside the horizon are connected, in the presence of back reaction, with (\ref{eq15}). In our case, to conveniently address the entropy spectrum, we rewrite the connection (\ref{eq15}) between the relevant modes as
\begin{eqnarray}
\Phi_{\textrm{in}}^L&=&\Phi_{\textrm{out}}^L,\nonumber\\
\Phi_{\textrm{in}}^R&=&e^{-\frac{1}{2}\beta_{\textrm{eff}}E}\Phi_{\textrm{out}}^R,\label{eq22}
\end{eqnarray}
where we have used the relations between (\ref{eq13}) and (\ref{eq14}), and $\beta_{\textrm{eff}}$ is the effective (inverse) temperature, whose value has been addressed in (\ref{eq20}). From (\ref{eq22}), we find the left mode is trapped inside the horizon, while the right mode can tunnel across the horizon to be observed at the asymptotic infinity. Thus, the average value of energy $E$ is given by
\begin{eqnarray}
\langle E\rangle=\frac{\int_0^\infty (\Phi_{\textrm{in}}^R)^\ast E \Phi_{\textrm{in}}^R dE }{\int_0^\infty (\Phi_{\textrm{in}}^R)^\ast  \Phi_{\textrm{in}}^R dE}&=&\frac{\int_0^\infty e^{-\frac{1}{2}\beta_{\textrm{eff}}E}(\Phi_{\textrm{out}}^R)^\ast E e^{-\frac{1}{2}\beta_{\textrm{eff}}E}\Phi_{\textrm{out}}^R dE }{\int_0^\infty e^{-\frac{1}{2}\beta_{\textrm{eff}}E}(\Phi_{\textrm{out}}^R)^\ast  e^{-\frac{1}{2}\beta_{\textrm{eff}}E}\Phi_{\textrm{out}}^R dE}\nonumber\\
&=& \frac{\int_0^\infty E e^{-\beta_{\textrm{eff}}E}dE}{\int_0^\infty  e^{-\beta_{\textrm{eff}}E}dE}\approx\beta_{\textrm{eff}}^{-1}. \label{eq23}
\end{eqnarray}
 In a similar way, the average squared energy of particles is detected by the asymptotic observer as
\begin{eqnarray}
\langle E^2\rangle=\frac{\int_0^\infty (\Phi_{\textrm{in}}^R)^\ast E^2 \Phi_{\textrm{in}}^R dE }{\int_0^\infty (\Phi_{\textrm{in}}^R)^\ast  \Phi_{\textrm{in}}^R dE}&=&\frac{\int_0^\infty e^{-\frac{1}{2}\beta_{\textrm{eff}}E}(\Phi_{\textrm{out}}^R)^\ast E^2 e^{-\frac{1}{2}\beta_{\textrm{eff}}E}\Phi_{\textrm{out}}^R dE }{\int_0^\infty e^{-\frac{1}{2}\beta_{\textrm{eff}}E}(\Phi_{\textrm{out}}^R)^\ast  e^{-\frac{1}{2}\beta_{\textrm{eff}}E}\Phi_{\textrm{out}}^R dE}\nonumber\\
&=& \frac{\int_0^\infty E^2 e^{-\beta_{\textrm{eff}}E}dE}{\int_0^\infty  e^{-\beta_{\textrm{eff}}E}dE}\approx2\beta_{\textrm{eff}}^{-2}. \label{eq24}
\end{eqnarray}
It should be noted that the results (\ref{eq23}) and (\ref{eq24}) are based on the assumption that the black hole has much more energy than the emitted particle ($M\gg E$). In this case, the inverse
temperature $\beta_{eff}$ can be approximately treated as a constant with the integration of the energy $E$.
Hence, the uncertainty in the detected energy $E$ is read off
\begin{equation}
\Delta E=\sqrt{\langle E^2\rangle-\langle E \rangle^2}=\beta_{\textrm{eff}}^{-1}=T_{\textrm{eff}}.\label{eq25}
\end{equation}
Here, $T_{\textrm{eff}}$ is nothing but the effective Hawking temperature.  This uncertainty can be treated as the lack of information in energy of the black hole during the particle emission. Also, in the information theory,
the entropy acts as the lack of information, so we can use the first law of thermodynamic as $\Delta E=T_{\textrm{eff}}\Delta S$ to  connect these two quantities. On the other hand, from (\ref{eq25}), we can easily find the uncertainty in the characteristic frequency of the tunneling mode as $\Delta f=\frac{\Delta E}{\hbar}=\frac{T_{\textrm{eff}}}{\hbar}$. Now, we have
\begin{equation}
S=\int \frac{\Delta E}{T_{\textrm{eff}}}=\frac{1}{\hbar}\int \frac{\Delta E}{\Delta f}.
\end{equation}
Then, according to the Bohr-Sommerfeld quantization rule, $\int \frac{\Delta E}{\Delta f}=n\hbar$,
we can immediately infer that the entropy is quantized and the entropy spectrum is given by
\begin{equation}
 S=n,\label{eq26}
\end{equation}
where $n=1,2,3,\cdot\cdot\cdot\cdot\cdot\cdot$. From (\ref{eq26}), we find $\Delta S=(n+1)-n=1$, this result showing that the entropy at the horizon of the black hole is quantized in units of the identity. In summary, the entropy spectrum in the presence of back reaction is $S_n=n$, sharing the same form as (\ref{eq10}) where the effects of back reaction are not incorporated,  which is to say, the entropy quantum is independent of back reaction. This is an interesting result.

\section{Conclusions and Discussions}\label{sec4}
In this paper we suitably adapt the \emph{reformulated} tunneling framework \cite{c1,c2,BRM,RBE,JHC} so that it can well accommodate the effects of back reaction to produce the Hawking emission spectrum and entropy spectroscopy.
Our derivation is universal and intriguing. Now, we summarize our conclusions and discussions as follows:

i) In the reformulated tunneling framework, the Hawking emission spectrum with back reaction deviates from the precisely thermal one, but is related to the difference of the Bekenstein-Hawking entropy before and after the particle emission. Most importantly, the effect of back reaction also implies a unitary theory underlying the process of black hole evaporation. In spite of these progress, it is not at all clear as to how the information might actually be preserved. When emitting two particles $E_1$ and $E_2$, and one particle with their combined energies $E_1 +E_2$, we find the probability of emitting two particles is
exactly the same as the probability of emitting just a single particle of the same total
energy, so there is no late-time correlation. That is to say, the form of the correction due to back reaction is
not sufficient by itself to relay information. Note that our above analysis is only described in the semiclassical framework. However, the true tunneling process should go beyond the semiclassical approximation. As a result, the black hole entropy would undergo higher-order quantum corrections. For the Schwarzschild black hole, the entropy is quantum-corrected as
\begin{eqnarray}
S_{\textrm{BH}}^\textrm{q}&=&\frac{4\pi M^2}{\hbar}+8\pi \sum_i\alpha_i\hbar^{(i-1)}\int M^{(1-2i)}d M+\textrm{constant}.
\end{eqnarray}
Here, the first term is the semiclassical entropy of the Schwarzschild black hole, which is related to one-quarter of its horizon area, also called as the Bekenstein-Hawking entropy. The other term come from higher-order quantum corrections. Now, another question arises as to whether the black hole information
is preserved by additionally incorporating the effects of higher-order quantum corrections. On the premise that the total tunneling rate (\ref{eqA3}) is preserved even in the presence of higher-order quantum corrections, we then have $ P^\textrm{q}=e^{\Delta S_{\textrm{BH}}^\textrm{q}}=e^{S_{\textrm{BH}}^\textrm{q}(M-E)-S_{\textrm{BH}}^\textrm{q}(M)}$. After just a few more
iterations, it is not difficult to convince oneself that $\ln[P^\textrm{q}(E_1)P^\textrm{q}(E_2)]=\ln[P^\textrm{q}(E_1+E_2)]$, meaning the probability of emitting two particles of energy $E_1$ and $E_2$ is
precisely the same as the probability of emitting just a single particle of the same total
energy $E_1+E_2$. Hence, the inclusion of such
corrections is not sufficient to account for the late-time correlations. That is to say, even
including the effects of back reaction and higher-order quantum corrections, the tunneling formalism can still
not provide a mechanism for preserving the black hole information. The best way to understand these outcomes is: Although the tunneling paradigm and the quantum-corrected entropy can be viewed as manifestations of quantum-gravitational principles, the tunneling program, as it currently stands, is formulated at only the level of semi-classical gravity. To better understand these issues, we might resort to quantum gravitational treatment.

ii) In the presence of back reaction, the entropy spectrum shares the same form $S_n=n$ as that without such effect. That means the entropy quantum is independent of the effects of back reaction. Also, in Ref. \cite{JHC} we observe that the entropy spectrum describing in the tunneling picture is independent of the high-order quantum corrections. Hence, we conclude that the entropy quantum is universal in the sense that it is not only independent of back reaction, but also independent of higher-order quantum corrections. This is an intriguing result for quantum-gravitational principles.

\appendix

\section{The unitary process for black hole emission}

In the appendix, we will confirm that, in the presence of back reaction, black hole emission process is consistent with the underlying unitary theory. Due to the effect of back reaction, the left and right moving modes defined inside and outside the horizon are connected with (\ref{eq15}) in the reformulated tunneling framework.
For the left modes, the observer staying outside the horizon would find that it can classically fall into the center of black hole. So the tunneling rate for the left modes should be expected to be unity. This can be easily verified by
\begin{equation}
P^{L}=|\Phi_{\textrm{in}}^{L}|^2=|\Phi_{\textrm{out}}^{L}|^2=1.
\end{equation}
Proceeding in the similar way, for the right modes, its probability, as seen by an external observer, is
\begin{equation}
P^{R}=|\Phi_{\textrm{in}}^{R)}|^2=e^{\Delta S_{\textrm{BH}}}|\Phi_{\textrm{out}}^{R}|^2=e^{\Delta S_{\textrm{BH}}}.
\end{equation}
The total tunneling rate is then given by
\begin{equation}
P=\frac{P^{R}}{P^{L}}=e^{\Delta S_{\textrm{BH}}}. \label{eqA3}
 \end{equation}
 Note that this observation is exactly in agreement with the underlying unitary theory. In quantum mechanics, the transition probability of a physical system from the initial states to the final states, is expressible as
\begin{equation}
P(i\rightarrow f)=|M_{fi}|^2\cdot (\textrm{phase space factor}),
\end{equation}
where the first term on the right side is the square of the amplitude for the process. For black hole,
the phase space factor is given by
\begin{equation}
\textrm{phase space factor}=\frac{N_f}{N_i}=\frac{e^{S_{\textrm{final}}}}{e^{S_{\textrm{initial}}}}=e^{\Delta S}.
\end{equation}
Here, $N_f$ and $N_i$ are, respectively, the number of final and initial states for the system, whose values for black hole can be precisely determined by the final and initial Bekenstein-Hawking entropy. Just such an outcome complies perfectly with
statistical considerations and thus implies a unitary evolving system.

\section*{Acknowledgments}
This work is supported by the Natural Science Foundation of China under Grant No.
11005086.

\end{document}